# Ex Luna, Scientia: The Lunar Occultation eXplorer (LOX)

*An Activity, Project, and Statement of the Profession (APC) White Paper*

Submitted by:
**Dr. Richard S. Miller**
The Johns Hopkins University Applied Physics Laboratory
Mail Stop 200-W230
11101 Johns Hopkins Road
Laurel, MD 20723-6099
240-592-1576
richard.s.miller@jhuapl.edu

On behalf of the LOX Collaboration:
**M. Ajello[1], J.F. Beacom[2], P.F. Bloser[3], A. Burrows[4], C.L. Fryer[3],
J.O. Goldsten[5], D.H. Hartmann[1], P. Hoeflich[6], A. Hungerford[3],
D.J. Lawrence[5], M.D. Leising[1], P. Milne[7], P.N. Peplowski[5],
F. Shirazi[1], T. Sukhbold[2], L.-S. The[1], Z. Yokley[5], C.A. Young[8]**

[1]Clemson University
[2]The Ohio State University
[3]Los Alamos National Laboratory
[4]Princeton University
[5]The Johns Hopkins University Applied Physics Laboratory
[6]Florida State University
[7]University of Arizona - Steward Observatory
[8]National Aeronautics and Space Administration

# 1. Introduction

*LOX* is a lunar-orbiting nuclear astrophysics mission that will probe the Cosmos at MeV energies (Ex. 1). It is guided by open questions regarding thermonuclear, or Type-Ia, supernovae (SNeIa) and will resolve the enigma of these inherently radioactive objects by enabling ***a systematic survey of SNeIa at gamma-ray energies*** for the first time.

Astronomical investigations from lunar orbit afford new opportunities to advance our understanding of the cosmos. The foundation of *LOX* is the lunar occultation technique (LOT), an observational approach well suited to the all-sky monitoring demands of supernova investigations and time-domain astronomy. Its inherently wide-field-of-view and continuous all-sky monitoring provides an innovative way of addressing decadal survey questions at MeV energies (0.1–10 MeV).

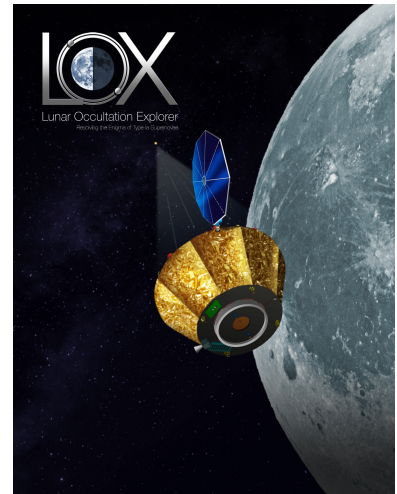

**Exhibit 1**. Artist's impression of the Lunar Occultation eXplorer (*LOX*).

The *LOX* approach achieves high sensitivity with a simple instrument design that eliminates the need for complex, position-sensitive detectors, kinematic event reconstruction, masks, or other insensitive detector mass, while also mitigating technology development, implementation complexity, and their associated costs (e.g., [2, 4, 45]). ***LOX can be realized within existing programs, like Explorer***.

The tremendous potential of MeV gamma-ray astronomy in general, and supernova radioactivity measurements in particular, is currently unrealized for the simple reason that instrument sensitivity has been inadequate; in fact, from 1980 until today, sensitivity has improved by only a factor of ten. This contrasts markedly with advances in soft X-ray, hard X-ray, GeV gamma-ray, and TeV gamma-ray astrophysics. By employing a paradigm changing observing methodology, ***LOX will utilize the Moon as a platform for astrophysics and change this status quo.***

LOX will address high-value SNeIa science by characterizing the spectral evolution of their primary nuclear gamma-ray emissions to probe fundamental processes and diversity, and expose new connections between matter–energy life cycles within galaxies. The anticipated opportunity for an Explorer-class mission is timely since it coincides with operation of the Large Synoptic Survey Telescope (LSST, www.lsst.org) and NASA's Wide-Field Infrared Survey Telescope (WFIRST, wfirst.gsfc.nasa.gov) in the mid-2020s to provide enhanced SNeIa-related context measurements and science return.

> *LOX* Addresses High-Priority, High-Value Science
>
> - *LOX* enables systematic studies of SNeIa and their progenitors
> - *LOX* will quantify SNeIa population systematics
> - *LOX* will reveal mass-energy lifecycles within galaxies
> - *LOX* will enable near-continuous all-sky monitoring supporting multiple topics in MeV astrophysics
> - *LOX* offers new capabilities in time-domain astronomy and surveys

## 2. Key Science Goals and Objectives

Thermonuclear supernovae are deeply connected to topics throughout astrophysics and cosmology. They have synthesized the majority of iron in the universe, supplied much of the energy input to the interstellar medium (ISM), and may have also revealed to us the fate of the universe. They are, however, an enigma. ***What are their progenitor systems? Why do they explode? How diverse is their population?*** The answers to these fundamental questions will have far-reaching impact, and they have motivated several unresolved Astro2010 Decadal Survey findings (Ex. 2) and multiple Astro2020 Decadal Survey science white papers [1, 6, 26, 34, 44, 50, 52, 56, 64, 66].

The clues that reveal the identity of the SNeIa progenitor systems, and the physical conditions that govern their explosions, remain elusive. The consensus is that SNeIa result from degenerate carbon-oxygen (CO) white dwarfs (WD) undergoing thermonuclear runaway [22, 23]. But a standard model picture is difficult to produce from ultraviolet, optical, and infrared (UVOIR) based trends, even for a single progenitor channel [43]. Information loss is a natural consequence of the nuclear radiation reprocessing responsible for the UVOIR signatures [8]; complex interstellar and circumstellar environments also affect this secondary emission. Therefore, despite considerable theoretical modeling efforts (e.g. [5, 14, 15, 41, 60, 61]), the final verdict on SNeIa progenitors and their explosion mechanisms has not yet been reached. It is possible that multiple evolutionary pathways contribute to the observed SNeIa population.

| Astro2010 Decadal Survey Findings | Top-Level Science Goals | Top-Level Science Objectives |
|---|---|---|
| What are the progenitors of SNeIa and how do they explode? (SSE2) | Characterize the Spectral Evolution of Type-Ia Supernovae | Parameterization of Type-Ia Gamma-Ray Light Curves |
| Why is the Universe accelerating? (CFP2) How do stars form? (PSF1) | Quantify the Diversity of Type-Ia Supernovae | Population Studies of Type-Ia Supernovae Spectral Evolution, Including Identification of Subclasses |
| What controls the mass–energy–chemical cycles within galaxies? (GAN2) "Areas of Unusual Discovery Potential": Time-Domain Astronomy & Surveys | Probe the Thermonuclear Physics & Standardization of Type-Ia Supernovae | Perform Census of Type-Ia Supernovae Progenitor Subclasses |
| | | Perform Census of Type-Ia Supernovae Environments |
| | | Connection to Type-Ia Supernovae UVOIR Diagnostics |

**Exhibit 2. Summary of LOX science goals & objectives.** Also shown are related Astro2010 Decadal Survey findings.

Gamma rays have been understood to be an excellent diagnostic of SNeIa for nearly five decades (e.g., [9, 13]). In contrast to UVOIR emission, the escaping nuclear radiation is a <u>direct</u> consequence of the nuclear physics that governs the structure and dynamics of these objects and is encoded with critical fundamental evidence that can travel great distances unscathed. The energy and time dependence of emergent gamma-rays provides unique and powerful insights into SNeIa and are the obvious choice as a probe to resolve the SNeIa enigma. ***Only measurements of this radioactivity—the ashes of nuclear burning—will reveal fundamental details of thermonuclear supernovae***.

Exhibits 3 and 4 are an illustrative example of the information content and diagnostic capabilities of nuclear gamma-ray measurements. These figures shows the evolution of nuclear gamma-ray emission for two viable SNeIa progenitor models: a Chandrasekhar mass detonation [53] and the merger of two white dwarfs [42]. At early epochs following the explosion, the magnitude of low-energy emission is significantly different for the two models. This is an indicator of radioactive elements buried beneath different amounts of material and a consequence of their distinct internal structures. At late epochs, when the ejecta are optically thin and gamma rays can escape freely, the spectra are dominated by the nuclear lines and are similar in intensity, reflecting the identical amount of $^{56}$Ni (and hence $^{56}$Co) produced in these two particular models.

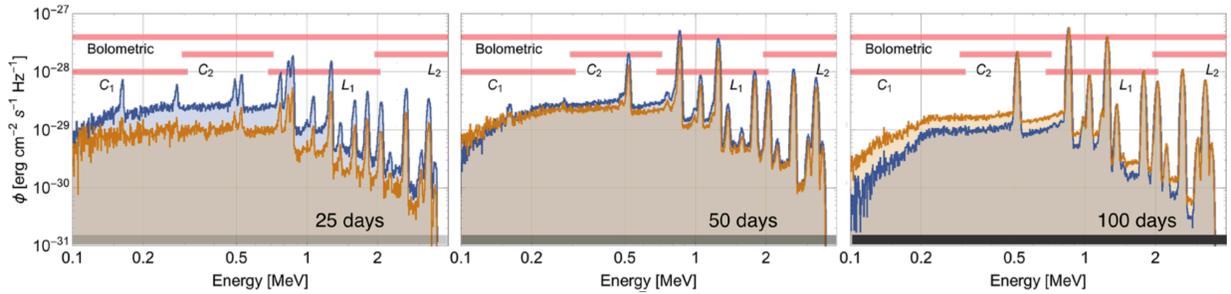

**Exhibit 3. Spectral evolution encodes information about SNeIa progenitor systems.** (left to right) Simulated emission spectra at 25, 50, and 100 days post-explosion for two representative SNeIa progenitor models [42, 53]. Line widths are due to Doppler broadening. Horizontal red lines denote preliminary definitions for continuum energy bands that, in addition to individual gamma-ray lines, contain information about elemental origin and transport processes.

Gamma-ray energy bands (i.e., continuum energy bands) also contain information about elemental origin and transport processes (e.g., [54, 55]) relevant for SNeIa studies. This contrasts with the approach requiring high-spectral resolution that seeks to characterize the discrete energy of nuclear decay lines (e.g., [24, 49]). The benefits of this approach include an increase in sensitivity due to additional gamma-ray flux, while also preserving the information of interest. Spectral evolution and intensity changes are therefore also be summarized in energy-band light curves, which characterize the spectral evolution of SNeIa. One notional set of definitions for SNeIa energy bands is given in Ex. 3.

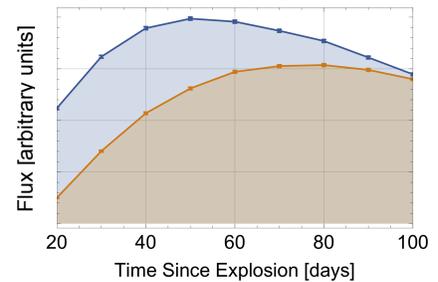

Diagnostics supporting SNeIa progenitor classification and the extraction of physical parameters can be obtained directly from the gamma-ray spectra and their corresponding temporal evolution in the form of light curves (e.g., [38, 54, 55]). The rapid expansion of the ejecta exposes successively deeper regions, allowing time-dependent measurements of gamma-ray emission to probe the abundance and distribution of radioactivity within the supernovae.

**Exhibit 4. Illustrative gamma-ray light curves.** Light curves (bolometric band, 0.1-4 MeV) for the two progenitors in Ex. 3.

After ~40 years of theoretical predictions and experimental efforts, only a single SNeIa (SN2014J) has been detected at MeV energies [7, 11, 24]. Despite its proximity (~3.5 Mpc), this source was at the limiting sensitivity of existing instruments and highlights ongoing limitations in survey performance (sensitivity, field-of-view, survey cadence, etc) intrinsic to the MeV gamma-ray missions deployed to date. Although emergent gamma-rays were monitored for ~100 days, efforts to compare the observed light curves to physical models (e.g. [57]) are limited by statistics (e.g. [12]). With a 100-fold improvement in sensitivity, *LOX* will revolutionize nuclear astrophysics and enable transformational population studies of SNeIa.

## 2.1. Mission Overview

Temporal modulation is the foundation of the LOT, with the Moon serving as a natural occulting disk to generate the required modulation via repeated eclipses of astronomical sources. Source detection, localization, flux characterization, and long-duration monitoring are based on time series analysis of occultation events, detailed knowledge of the lunar and spacecraft ephemerides, and a rigorous statistical framework. The dense lunar regolith, absence of an appreciable atmosphere or magnetosphere, and well-characterized radiation environment enhance performance over terrestrial implementations. Lunar occultation enables uniform and continuous monitoring of the sky at full sensitivity.

A celestial target is occulted by the lunar disk when the relative orientation between it, the Moon, and a detector meets a well-defined geometric condition. *LOX* uses <u>ensembles</u> of event occultations to detect and characterize astrophysical gamma-ray sources. Flux sensitivity, spectral resolution, field of view (FoV), and source localization are governed by implementation parameters such as spectrometer size, detector type, orbit altitude, and spectrum integration times (i.e., mission-level rather than technology-driven solutions). Baseline mission performance parameters are provided in Ex. 5. The *LOX* concept is highly scalable, limited only by resource constraints such as mass and power.

| Mission Summary | |
|---|---|
| Energy Range | 0.1-10 MeV |
| Prime Mission Duration | 3 years |
| Operational Duty Cycle | ≥95% |
| Orbit | Lunar |
| Orbit Altitude | 1500×40000 km (notional) |
| Spectrum Acquisition Cadence | <10 sec |
| Spectral Resolution | <9.5% FWHM @ 0.662 MeV |
| Source Localization | ~1 arcmin |
| Payload Mass (baseline) | 877 kg |
| Total Dry Mass | 1340 kg |
| Instrument Power | <100 W |
| Data Rate | ~20 kbits/s |
| Nadir Tracking Requirement | ≤1° |
| Science Data Return | ≥4 Tb |
| Mission Classification | Category 2, Class C |

**Exhibit 5**. Table of *LOX* mission parameters. Based on the *LOX* baseline configuration.

*Simplicity is a hallmark of the LOX concept*. It requires only a non-imaging spectrometer in lunar orbit. *LOX* will place a large-area gamma-ray spectrometer array into lunar orbit and continuously acquire time-resolved broadband spectra. The instrument will be pointed to the nadir (i.e., toward the center of the Moon), and sources will repeatedly rise and set along the lunar limbs. As *LOX*'s orbit evolves with respect to the celestial sphere, large swaths of the sky are surveyed. Operations are also simple because there are no slewing or onboard data-processing requirements. Data-analysis protocols are based on flexible and established time-series analyses of acquired spectra [30, 35, 37] which, although spectra are acquired continuously, can be divided into (artificial) observing periods to facilitate the monitor and characterization of evolving source light curves.

*LOX's location in lunar orbit provides many advantages over traditional, Earth-orbiting gamma-ray observatories*. The well-characterized, slowly changing lunar background environment enables the mitigation of background systematics, the benefits of extended observing periods, and the statistical advantages of using occultation event ensembles [39].

Additionally, the gamma-ray backgrounds from the lunar surface provide an in-situ calibration source that reduces associated systematics to the level of few percent [30]. In contrast, Earth-orbiting observatories contend with dynamic and complex background environments that change on orbital timescales and are not easily characterized (e.g., [21]).

Target source locations can be identified uniquely in the sky using the *LOX* time-series data (Ex. 6). This uniqueness condition is guaranteed by instantaneous occultation onset times (for point sources), which are themselves ensured by the Moon's high density and lack of atmosphere. This instantaneous occultation cannot be achieved from Earth orbit because of its atmosphere [21, 48]. Precisely resolving occultation times translates directly into source-localization performance and source-confusion mitigation.

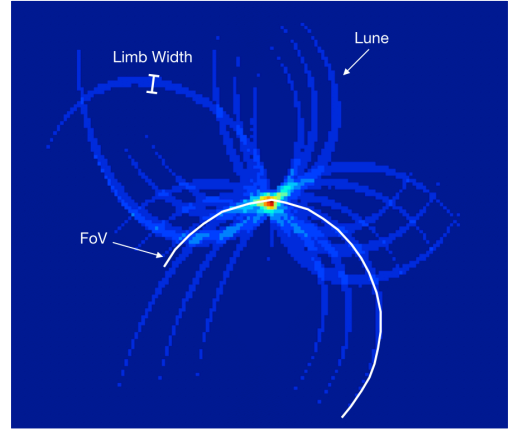

**Exhibit 6**. **Example of LOX source localization**. The source sits at locus of lunar limbs (lunes) projected onto the celestial sphere. Field-of-view (FoV) and lune width are defined by the S/C orbit.

All aspects of the LOT— including source-analysis methodologies [35, 36, 37, 39, 40], extended operation of instrument components in lunar orbit [16, 17, 28, 29], and the mitigation of systematics driven by the lunar background environment [30, 35]—have been validated from lunar orbit with previous deployed assets.

## 2.2. Performance Estimate

Achievable continuum sensitivity for the LOX baseline mission is shown in Ex. 7, with selected performance parameters tabulated in Ex. 5. The desire to identify a large population of SNeIa (>100 for a 3-year mission) necessitates large sky coverage and excellent sensitivity for supernovae to a horizon beyond 100 Mpc. Uniformity of exposure, and therefore sensitivity, is a defining feature of *LOX* and has not been generally achieved using other observing methods [27, 49, 51]. *LOX*'s continuum sensitivity is achieved over >85% of the sky for each standard observing period ($10^6$ s); complete sky coverage at full sensitivity is achieved on a timescale of several weeks for the lunar orbits currently under consideration.

It is important to note that the LOT methodology differs in important ways from Earth occultation approaches [39]. Event ensembles (i.e. sets of occultation intervals) enhance statistical power and are uniquely enabled by the presence of a stable lunar background environment. This contrasts markedly with Earth-occultation endeavors, which are limited to the analysis of individual occultation events to mitigate systematics produced by the dynamic backgrounds present in Earth-orbit. Another key advancement is the use of an advanced statistical framework to obtain independent, and data-derived, source and background flux estimates for an uncollimated detector without extensive modeling. LOT-based flux sensitivity utilizes all acquisition intervals within an observing period, not only those in which a source is known to rise or set, since it's the template of occultations that is relevant to a given source location. An intrinsic mechanism for identifying and monitoring systematics is also afforded by this approach.

## 2.3. Relevance and Impact

One of the most stunning discoveries of the last two decades was the realization that the Universe is undergoing an accelerated expansion, suggesting a new form of energy — Dark Energy — constitutes ~70% of the energy budget of the Universe. This discovery was based on the assumption that UVOIR SNeIa light curves can be used as quasi-standard candles with an accuracy of about 15%. The quest for the nature of Dark Energy, however, requires accuracies closer to 1%, a level that drives many current and upcoming surveys and satellites including ASASSN, CSP, PTF, ZTF, LSST, HST, WFIRST and JWST.

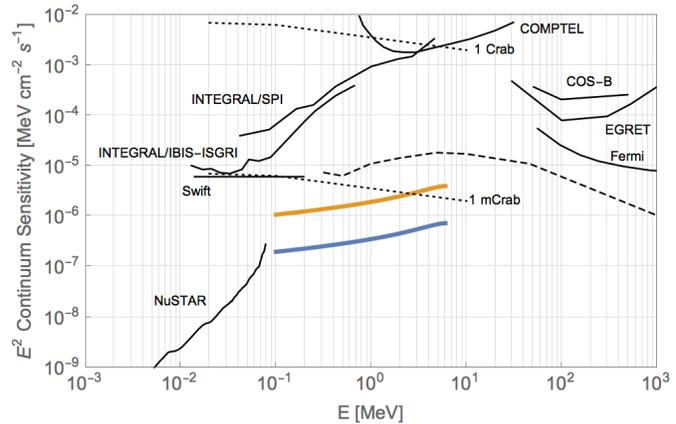

**Figure 7**. **Continuum sensitivity of *LOX* baseline mission**. Sensitivity is 3σ detection significance, ΔE/E=1. *LOX* exposure is $10^6$ seconds (gold) and 1 year (blue) of elapsed time; all others are $10^6$ seconds on-source. Predicted sensitivity of other proposed mission concepts [2, 45] are also shown (dashed).

Precision cosmology, including probes of the Dark Energy equation of state, requires standardizing SNeIa below the 1% level. Some of the primary challenges affecting the standard UVOIR-based techniques are the dust properties of the interstellar medium (ISM) and complexity inherent in the interpretation and modeling of observed spectra. In addition, the source of these low-energy photons is reprocessed energy from the nuclear decay (i.e. the fundamental gamma-rays) and their interpretation depends on complex simulations of radiation transport and its coupling to atomic-level populations, ionization states in a hot plasma, and/or empirical methods that use large samples but are subject to potential effects of redshift evolution. In contrast, gamma-rays measure the energy-input directly, the gamma-ray opacities do not depend on temperature or ionization because the plasma is cool compared to the photon energy, and gamma-rays are hardly affected by the ISM.

Significant progress has been made in SNeIa-based cosmology (e.g. [46, 47]), resulting in a new mystery: an apparent disagreement in a determination of the Hubble parameter [47, 52] between measurements made at early (cosmic microwave background) and late (SNeIa) times in the universe's evolution. While several exotic explanations have been proposed to resolve this mystery, an astrophysical explanation based on the (unknown) SNeIa systematics and/or (unquantified) population diversity is a viable alternative.

Finally, we note that the *LOX* performance parameters that enable transformational SNeIa science also enable us to address a breadth of additional science topics in astrophysics [1, 10, 18, 19, 20, 25, 31, 32, 33, 34, 50, 56, 58, 59, 63, 65].

## 3. Technical Overview

The *LOX* mission concept is based on simplicity of operation and high-heritage (low-risk) components. Mission design, flight hardware and software, ground systems, and the operations strategy, satisfy mission requirements with a low-risk solution that yields appropriate resource margins, uses simple interfaces and mature technologies, and draws upon lessons learned and extensive heritage from many missions.

> **LOX Leverages Past Experience in Lunar Gamma-Ray Spectroscopy Applications**
>
> - Single instrument payload, simple operation
> - Detector components are high-heritage and have extended operational history in lunar orbit
> - BAGEL instrument design is highly scalable and fault-tolerant
> - LOX data processing builds on >20 years of experience with lunar and other planet-orbiting GRSs

*LOX requires only standard launch services*. The *LOX* trajectory consists of a 3-month transfer to the Moon followed by 3 years in orbit. The transfer takes advantage of the weak stability area that occurs between the Sun and Earth and poses several advantages over a direct Earth–Moon transfer. Most importantly, it maximizes the amount of mass delivered into lunar orbit. Lunar orbit insertion (LOI) places the SC into an orbit specifically designed such that the periapsis and apoapsis altitudes remain stable over time, requiring no maintenance maneuvers. The orbit stability provides for long-duration science operations while requiring minimal consumables. The design has placeholders reserved for disposal requirements to be finalized upon mission selection.

## 3.1. Instrumentation

*LOX* adopts a large-area array of scintillator-based detectors as its single instrument. The primary detector material is bismuth germanate ($Bi_4Ge_3O_{12}$, BGO), which provides the highest gamma-ray absorption efficiency of available scintillators. BGO is well suited for spaceflight applications because it is very rugged, non-hygroscopic (hermetic encapsulation not required), and very dense (compact), which is why it was also the scintillator of choice for the Lunar Prospector, NEAR Shoemaker, and Dawn GRS sensors, as well as the Fermi Gamma-Ray Burst Monitor (GBM) [3].

The BGO Array for Gamma-Ray Energy Logging (BAGEL) is composed of an array of individual detector modules (Ex. 8), and a set of common electronics boxes sub-divided into Data Processing Units (DPU) for fault tolerance. Each BAGEL module is self-contained and operates independently; spectra are accumulated on time intervals common to the entire array. The detector array is configured to minimize shadowing effects from neighboring modules and to mitigate induced backgrounds. Although other detector materials have superior spectral resolution, BGO (Ex. 5) is sufficient to meet all science requirements and has significant flight heritage. Our mission baseline employs an array of 100 BAGEL modules (Ex. 8), with the final number limited only by available launch mass rather than cost.

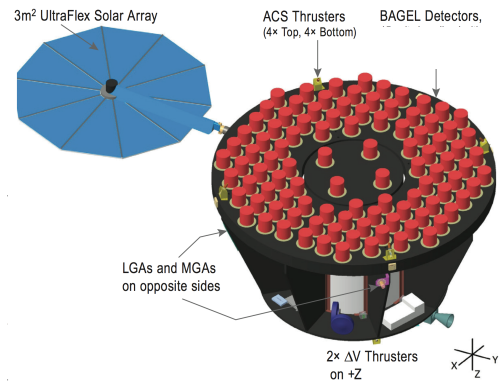

**Exhibit 8**. **LOX spacecraft concept.** The BAGEL instrument (red) is an array of high-heritage gamma-ray spectrometers.

Each module is configured and operates as a *phoswich* sensor. The heart of each module is a 127 mm (5") diameter by 127 mm (5") long BGO crystal. Surrounding the BGO is a thin (~6 mm) well-shaped BC-408 plastic scintillator, which is primarily sensitive to charged particles. An outer layer of diffuse reflector material ensures uniform light collection. The scintillators are optically coupled and read out by a single Hamamatsu R877 PMT, the same model used in the Fermi GBM. The plastic scintillator acts as an anti-coincidence shield (ACS), vetoing cosmic ray events that would otherwise introduce a background that would lower the signal-to-background ratio of the system. Pulse processing techniques separate the signals from the two dissimilar scintillators based on their distinct differences in rise times, and existing performance measures exceed mission requirements. This phoswich approach minimizes implementation complexity (e.g. number of electronics channels) and instrument mass.

### 3.2. Operations

The science observing profile for *LOX* is remarkably simple and tolerant. The selected orbit will balance competing criteria by maximizing the occultation field of view while minimizing backgrounds from diffuse emission and the dominant lunar albedo. Candidate lunar orbits have been identified that require little to no maintenance. *LOX* operates continuously, producing spectra at a steady cadence. There are no time-critical events. Nadir-pointing requirements are modest, with knowledge more important than accuracy. Any off-pointing times (e.g., safing) are expected to account for <5% of operations.

The *LOX* ground system architecture and MOps concept leverages heritage in providing an efficient implementation that fully meets the mission functional requirements. The Johns Hopkins University Applied Physics Laboratory (APL) Mission Operations engineers and analysts have extensive space experience from NEAR, MESSENGER, STEREO, New Horizons, Van Allen Probes, and Parker Solar Probe. As with all APL missions, the *LOX* Mission Operations Team is established early in the life cycle and works closely with development leads to provide insights that streamline end-to-end operation and reduce operational cost and risks. *LOX* operations will be conducted from one of APL's state-of- the-art Mission Operations Centers.

## 4. Technology Drivers

*LOX* combines established detector technology with heritage electronics to minimize both implementation and operational risk. ***No new technologies are required***. Technology Readiness Levels (TRL) for all sub-systems are TRL $\geq$ 6, with many at TRL 9. The one exception is the BAGEL module which is currently assigned a TRL of 5, although all individual detector components are flight-proven (TRL 9). Performance testing in a near-space environment will raise the BAGEL module TRL to 6 using a high-altitude balloon flight planned in 2020, i.e. prior to the *LOX* proposal submission. Laboratory tests performed to date have demonstrated the implementation approach will meet all science requirements, with ample margin, while staying within the anticipated payload resources and cost constraints of the Explorer program.

## 5. Organization, Partnerships, and Current Status

### 5.1. Organization & Partnerships

*LOX* is organized by employing a philosophy of simplicity and efficiency. The *LOX* project organization is simple, flat, and streamlined. The *LOX* management team is composed of a small group of leading scientists from the astrophysics and planetary science communities and space system engineers with an impressive record of managing NASA planetary missions. The team has proven leadership qualities, extensive experience, and access to modern facilities and resources to ensure (a) execution of a focused mission meeting all science requirements, (b) adherence to NASA certified engineering and mission assurance procedures, (c) reduction of all risks to an acceptable level, and (d) mission compliance within approved cost and schedule agreements. The *LOX* key management team includes the Principal Investigator (PI), Program Manager (PM), Project Scientist (PS), and Mission System Engineer (MSE).

APL is the primary implementing institution for the LOX mission, and the entire implementation team is colocated at APL, facilitating communication and making it easy to quickly assemble team members to address issues. This institution has been involved in the national space program since 1959, has successfully launched 65 operational satellites, and developed >200 science instruments flown on a variety of NASA missions. APL also has considerable experience in leadership roles as the designated primary implementing institution for NASA PI-led missions including MESSENGER, New Horizons, Van Allen Probes, Parker Solar Probe, and the recently selected Dragonfly. A natural consequence of this engagement is that APL has evolved a highly experienced staff of scientists and engineers, advanced analytical and modeling skills, and modern test facilities that provide excellent end-to-end capabilities.

| Institution | Type | Role |
|---|---|---|
| Johns Hopkins University Applied Physics Laboratory | University Affiliated Research Center | Implementing Institution, Science Team |
| University of Arizona - Steward Observatory | University | Science Team |
| Clemson University | University | Science Team |
| Florida State University | University | Science Team |
| Goddard Space Flight Center | NASA Center | Science Team |
| Los Alamos National Laboratory | National Laboratory | Science Team |
| The Ohio State University | University | Science Team |
| Princeton University | University | Science Team |

**Exhibit 9**. *LOX* mission participating organizations.

The *LOX* PI is Dr. Richard S. Miller, Senior Professional Staff at the Johns Hopkins University Applied Physics Laboratory. To support the *LOX* science effort, Dr. Miller has organized a team of 18 Co-Investigator (Co-I) scientists recruited from the greater astrophysics community representing 8 leading research institutions (Ex. 9) and spanning multiple disciplines such as supernova modeling and phenomenology to nuclear and lunar science. To support PI oversight of the science team, Dr. Mark Leising, Clemson University, has been designated as the *LOX* Project Scientist.

### 5.2. Current Status

*LOX* combines established detector technology with heritage electronics in a straightforward way that minimizes both implementation and operational risk. The high-heritage design balances margins and reliability to maximize science return. These features, combined with lessons learned from, and development efforts since, our 2016 MIDEX submission mean that *LOX* is ready for the next MIDEX Announcement of Opportunity (AO). Data analysis methodologies currently exist at a level that provides high-confidence in science return estimates, and we anticipate that all mission subsystems will be at TRL ≥6 by summer 2020.

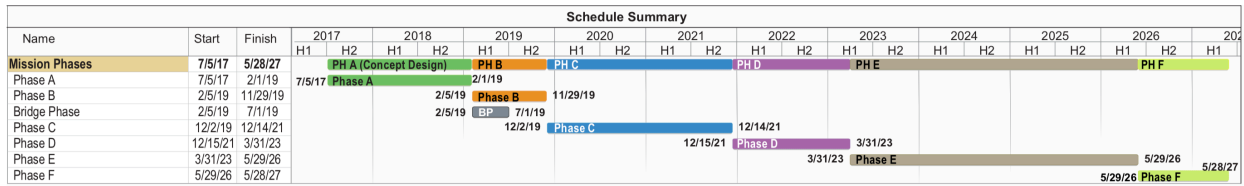

**Exhibit 10**. *LOX* **mission development schedule**. All mission phases are measured relative to the proposal submission, and is based on the *LOX* 2016 MIDEX submission.

## 6. Schedule

A *LOX* mission proposal will be submitted in response to the next MIDEX-class (astrophysics) Explorer Announcement of Opportunity, currently anticipated to be released in 2021. The development schedule from our 2016 MIDEX proposal is provided in Ex. 10; a similar timeline is anticipated in a future submission. Launch and operations will occur approximately 5 years after selection. The maturity of *LOX* subsystems mitigates schedule risk.

Submission of the Phase A concept study report is scheduled 9 months after selection, followed 2 months later by a site visit. Upon flight selection, a 10-month Phase B is scheduled, beginning with a 5-month Bridge Phase. The Bridge Phase provides a means to negotiate any early schedule risk-reduction subcontracts. Phase C/D extends 40 months through launch and a 1-month commissioning period. Phase E commences 3 months later with LOI. Science data collection will extend 3 years post-LOI. A 12-month Phase F is scheduled to accommodate final science data product generation and archiving.

## 7. Cost Estimates

A top-level cost summary is given in Ex. 11, based on detailed estimates made to support a 2016 MIDEX submission using the work breakdown structure (WBS) provided in the Explorer program library. Phase A–F Principal Investigator (PI)-managed mission cost (PIMMC) is $237M in fiscal year (FY) 2017 dollars, including appropriate reserves for all mission phases to provide sufficient protection against cost uncertainty and risk commensurate with a Class-C mission. The baseline LOX mission concept fits within the MIDEX Explorer-class of missions ($250M cost cap).

| Cost Summary | FY17 $M |
|---|---|
| PI-Managed Mission Cost + Reserves (Phases A-D) | 215 |
| PI-Managed Mission Cost + Reserves (Phases E-F) | 22 |
| Total Mission Cost | 237 |

**Exhibit 11**. *LOX* **mission cost summary**. Cost analysis was performed by The Johns Hopkins University Applied Physics Laboratory.

Robust and data-driven cost risk analysis gives confidence that LOX can be completed within the proposed cost. APL prepared the cost estimate using processes consistent with NASA's Cost Estimating Handbook and associated documents. All mission costs are estimated using a combination of parametric models, analogous historical costs, and bottom-up estimates (BUE). Independent, model-based cross-check estimates have been validated and calibrated for LOX-specific requirements using experience from recently completed NASA-sponsored APL projects.

# 8. References

(**BOLD** denotes Astro2020 Science White Papers).